\documentclass[
superscriptaddress,
twocolumn,
longbibliography,
pra,
aps,
10pt
]{revtex4-2}

\usepackage{graphicx}
\usepackage{amsmath,amsfonts,amssymb}
\usepackage{xcolor}
\usepackage{hyperref}
\hypersetup{colorlinks=true,allcolors=blue}
\usepackage[T1]{fontenc}
\usepackage{footnote}
\usepackage{siunitx}
\usepackage{layouts}
\usepackage{physics} 
\usepackage{dsfont}

\usepackage{tikz}
\usepackage{xcolor}
\usetikzlibrary{arrows.meta,calc,math,fit,positioning}
\definecolor{color_g}{HTML}{000000}
\definecolor{color_dm}{HTML}{4e6fef}
\definecolor{color_bm}{HTML}{4ebfef}
\definecolor{color_b}{HTML}{B4B4B4}
\definecolor{color_bp}{HTML}{B4B4B4}
\definecolor{color_dp}{HTML}{B4B4B4}
\definecolor{color_text}{HTML}{B4B4B4}

\renewcommand{\t}[1]{\textrm{#1}}
\renewcommand{\vec}[1]{\mathbf{#1}}

\begin{document}

\title{The optical Stark shift to control the dark exciton occupation of a quantum dot in a tilted magnetic field}
\newcommand{\UIBK}{Institute for Experimental Physics, University of Innsbruck, Innsbruck, Austria}
\newcommand{\WWU}{Institute of Solid State Theory, University of Münster, 48149 Münster, Germany}

\author{Miriam Neumann}
\affiliation{\WWU}
\author{Florian Kappe}
\affiliation{\UIBK}
\author{Thomas K. Bracht}
\affiliation{\WWU}
\author{Michael Cosacchi}
\author{Tim Seidelmann}
\author{Vollrath Martin Axt}
\affiliation{Theoretische Physik III, Universit{\"a}t Bayreuth, 95440 Bayreuth, Germany}
\author{Gregor Weihs}
\affiliation{\UIBK}
\author{Doris E. Reiter}
\affiliation{\WWU}
\date{\today}

\begin{abstract}
When a detuned and strong laser pulse acts on an optical transition, a Stark shift of the corresponding energies occurs. We analyze how this optical Stark effect can be used to prepare and control the dark exciton occupation in a semiconductor quantum dot. The coupling between the bright and dark exciton states is facilitated by an external magnetic field. Using sequences of laser pulses, we show how the dark exciton and different superposition states can be prepared. We give simple analytic formulas, which yield a good estimate for optimal preparation parameters. The preparation scheme is quite robust against the influence of acoustic phonons. We further discuss the experimental feasibility of the used Stark pulses. Giving a clear physical picture our results will stimulate the usage of dark excitons in schemes to generate photons from quantum dots. 
\end{abstract}

\maketitle

\section{Introduction}
The optical Stark effect relies on the application of a strong, detuned laser field. Applied to a few-level system as occurring in self-assembled semiconductor quantum dots (QDs), this results in an energy shift of the involved states \cite{jundt2008observation}. The optical Stark effect in QDs has been used in several ways: for ultrafast control of the exciton polarization \cite{Unold2004}, to reduce the fine-structure splitting between the exciton levels \cite{muller2009creating}, to induce the Autler-Towns splitting and for spin switching in QDs doped with a single magnetic ion \cite{le2011optical,reiter2012spin}, to enable the preparation of Fock states in QD-cavity systems \cite{Cosacchi2020b}, to enhance and suppress the tunneling rate between double-QD structures \cite{Tsukada1997}, and to control the energetic landscape of a charged QD in a spin-selective manner \cite{Wilkinson2019}. In other systems, the Stark effect has been used in GaAs quantum wells already in 1986 \cite{VonLehmen86,PhysRevLett.56.2748}, while recent applications of the Stark effect are used in the newly found 2D semiconductors \cite{cunningham2019resonant} as well as for the ultrafast control of exciton-polariton systems \cite{Panna2019}.

In this paper, we study how the optical Stark effect can be used to address the dark exciton in a QD, which is typically not coupled to optical transitions. We consider the dark exciton in the ground state manifold, where the spins of the constituent electron and hole are aligned parallel, while the excitons with anti-parallel spin can be optically addressed and are hence called bright excitons. While bright excitons in QDs are exploited in the usage of QDs as single-photon sources \cite{Ding2016,Somaschi2016,Schweickert2018,Hanschke2018,Cosacchi2019}, dark excitons can be used as auxiliary states to create more complex photon states, like photonic cluster states \cite{schwartz2016deterministic} or time-bin entangled states \cite{Jayakumar:2014aa,huber2016coherence}. Dark excitons in semiconductor materials gain more and more importance: in 2D semiconductors it was proposed to use dark excitons as sensors \cite{feierabend2017proposal} and in perovskite nanocrystals dark excitons can promote the photon emission rate \cite{tamarat2020dark}.

Experimentally implemented protocols to prepare the dark ground-state exciton rely on the decay of a higher energy biexcitonic state, heralded by photon detection \cite{poem2010accessing, schwartz2015deterministic, heindel2017accessing}. Higher-excited states addressed by a laser with a strong component along the beam axis can also relax into the dark ground-state exciton \cite{holtkemper2021dark}. Other protocols have been proposed using a single pulse relying either on adiabatic rapid passage \cite{luker2015direct} or on phonon-assisted state preparation \cite{luker2017phonon} in combination with a tilted magnetic field. A tilted magnetic field can also be used to extend the exciton-photon system of a QD in a cavity  \cite{jimenez2017dark,jimenez2020strong}. 

In the present scheme, we propose to use two pulses to prepare the dark exciton in a QD in a tilted magnetic field: an ultrafast $\pi$-pulse to excite the bright exciton followed by a rectangular pulse with softened edges to induce the optical Stark effect. The proposed scheme has the advantage of offering a clear picture of the underlying physics to prepare the dark exciton in QDs as well as being a very precise preparation method. We provide analytical equations for the parameters of the Stark pulse, thereby offering guidance on the usage of the Stark effect. With this method also superposition states can be easily prepared. We study the influence of acoustic phonons, which are known to often have a strong impact on exciton preparation schemes \cite{Ramsay2010a,luker2019review}, showing that they have only a marginal influence on this preparation scheme at low temperatures. We close by discussing the experimental feasibility of the proposed Stark pulses. 

\section{System and theoretical background}\label{sec:QD}
\subsection{System states}\label{sec:QDstates}
Strongly confined QDs host discrete excitons, with their ground state being strongly separated from higher-excited exciton states. Being composed of an electron with spin $S_e$ and a hole with (pseudo-)spin $S_\text{h}$, the spin combination $j=S_\text{e}+S_\text{h}$ determines the optical properties of the exciton. While the electron spin can be given by $S_\text{e} \in \{+\frac{1}{2}(\uparrow),-\frac{1}{2} (\downarrow)\}$, for holes we have to distinguish between light holes where $S_\text{lh} \in \{+\frac{1}{2},-\frac{1}{2} \}$ and heavy holes with $S_\text{hh} \in \{+\frac{3}{2} (\Uparrow),-\frac{3}{2}(\Downarrow) \}$. In typical QDs, the heavy holes constitute the hole ground state, while the light holes are energetically separated and can be neglected \cite{Reiter2014}, such that we use $S_\text{h}=S_\text{hh}$. We mention that for strongly strained QDs the light holes can constitute the (hole) ground states \cite{huo2014light}. 

The optical selection rules can be deduced by the total spin momentum of the exciton, which can be $j=\pm1$ or $j=\pm2$. Emitting (absorbing) a photon requires a change of $\Delta j = \pm 1$ for circular polarisation $\sigma^{\mp}$ ($\sigma^{\pm}$). This allows us to distinguish two species of excitons depending on their ability to couple to the electromagnetic field via pair recombination (creation): Bright excitons that have anti-parallel spin contributions $j = \mp S_\text{e} \pm S_\text{h} = \pm 1$ and dark excitons with parallel spin contributions $j = \pm S_\text{e} \pm S_\text{h} = \pm2$. 

\begin{equation*}
    \begin{aligned}
        &\ket*{b^{+}} = \ket*{\downarrow,\Uparrow} ~~~~ 
        &\ket*{b^{-}} = \ket*{\uparrow,\Downarrow} \,\,\,\, \\
        &\ket*{d^{+}} = \ket*{\uparrow,\Uparrow} ~~
        &\ket*{d^{-}} = \ket*{\downarrow,\Downarrow} \,.
    \end{aligned} 
\end{equation*}

The different excitons interact with each other via the Coulomb interaction, in particular, due to the exchange interaction. The short range exchange interaction results in a splitting between bright and dark excitons, while the long range exchange interaction yields an interaction between the bright excitons known as fine-structure splitting \cite{Bayer2002,gammon1996finestructure}. 

\subsection{Coupling to magnetic fields}
For our preparation scheme, we assume an external tilted magnetic field composed of an $x$- and $z$-component, which is applied to the QD.
The $z$-component of the magnetic field is helpful to adjust the energy levels in our preparation scheme as discussed below, while the $x$-component is responsible for the coupling between bright and dark exciton, resulting in the dark exciton gaining oscillator strength and becoming bright \cite{bayer2000spectroscopic}. In our model the in-plane magnetic field mediates a spin flip of the electron only, while the hole spin remains fixed. The latter is based on the assumption that a direct spin flip between the heavy hole states is forbidden. In principle a hole spin flip via the light hole states is possible, in particular in the presence of band mixing, however, this process is slow compared to the preparation protocol and is further prohibited by the out-of-plane magnetic field. 

Another possibility to couple bright and dark excitons is via valence band mixing, which is strong in highly asymmetric QDs. Due to the mixing, the dark exciton gains oscillator strength and can become visible \cite{schwartz2015deterministic,heindel2017accessing}.  In both of these cases, the dark exciton is not 100\% dark anymore and recombines radiatively, but typically on a much longer timescale than the bright exciton \cite{schwartz2015deterministic}. Also a magnetic dopant in the QD can act as an effective magnetic field on the exiton and result in spin flips \cite{besombes2004probing,reiter2009all}.

\subsection{Reduction to a three-level model}

\begin{figure}
    {\begin{tikzpicture}
	[
	arrow/.style={<->,shorten >=2pt, shorten <=2pt,>={Stealth[round]},semithick},
	arrow2/.style={<->,shorten >=0.5pt, shorten <=0.5pt,>={Stealth[round]},semithick},
	elecarrow/.style={->,thick},
	holearrow/.style={thick,double, double equal sign distance,-Implies},
	level/.style={<->,line width=2pt,>={Round Cap[length=0.5pt]}}]
	\tikzmath{
		\energy = 2.500;
		\Eb = -0.5000;
		\Edp = 0.2354;
		\Ebp = 0.7197;
		\Ebm = -0.1197;
		\Edm = -1.0354;
		\shift=0.5;
		\detuning=0.6;
	}
	\draw[level,color_g] (0.5,0) -- (-0.5,0) node[pos=.5,below=-1pt] {$\ket{g}$};
	\draw[level, color_bm] (0.5+\shift,\energy+\Ebm) -- (1.5+\shift,\energy+\Ebm) node[pos=.65,left=16pt,black] {$\ket{b^{-}}$};
	\draw [arrow, red,thick] (0.25,0) -- (0.75+\shift,\energy+\Ebm) 
	node [midway,right] {$\sigma^{-}$};
	\draw [arrow, color_bp] (-0.25,0) -- (-0.75-\shift,\energy+\Ebp) 
	node [midway,left] {$\sigma^{+}$};
	\draw[level, color_dm] (2+\shift,\energy+\Edm) -- (3+\shift,\energy+\Edm) node[pos=.5,below left,black] {$\ket{d^{-}}$};
	\draw[level, color_bp] (-0.5-\shift,\energy+\Ebp) -- (-1.5-\shift,\energy+\Ebp) node[pos=.6,above=-2pt] {$\ket{b^{+}}$};
	\draw[level, color_dp] (-2-\shift,\energy+\Edp) -- (-3-\shift,\energy+\Edp) node[pos=.5,above=-2pt] {$\ket{d^{+}}$};
	\draw[level, color_b] (0.5,\energy + \energy+\Eb) -- (-0.5,\energy + \energy+\Eb) node[pos=.5,above=-2pt] {$\ket{XX}$};
	\draw [arrow, color_bp] (0.75+\shift,\energy+\Ebm) -- (0.25,\energy + \energy+\Eb)
	node [midway,above right] {$\sigma^{+}$};
	\draw [arrow, color_bp] (-0.75-\shift,\energy+\Ebp) -- (-0.25,\energy + \energy+\Eb)
	node [midway,left=-3pt,above=-1pt] {$\sigma^{-}$};
	\draw [arrow] (2.25+\shift,\energy+\Edm) -- (1.25+\shift, \energy + \Ebm)
	node [midway,above=2pt,right] {$\propto B_x$};
	\draw [elecarrow,blue] (0.75+0.11+\shift,\energy+\Ebm-0.25) -- (0.75+0.11+\shift,\energy+\Ebm+0.25);
	\draw [holearrow, red] (1.25-0.11+\shift,\energy+\Ebm+0.25) -- (1.25-0.11+\shift,\energy+\Ebm-0.25);
	\draw [elecarrow,blue] (2.25+0.11+\shift,\energy+\Edm+0.25) -- (2.25+0.11+\shift,\energy+\Edm-0.25);
	\draw [holearrow, red] (2.75-0.11+\shift,\energy+\Edm+0.25) -- (2.75-0.11+\shift,\energy+\Edm-0.25);
	\draw [arrow2] (3.2+\shift,\energy+\Edm) -- (3.2+\shift,\energy+\Ebm)
    node [midway, right] {$\delta_{bd}$};
	\draw [thick,densely dashed] (3.2+\shift,\energy+\Edm) -- (3.7+\shift,\energy+\Edm);
	\draw [thick,densely dashed] (1.6+\shift,\energy+\Ebm) -- (3.7+\shift,\energy+\Ebm);
\end{tikzpicture}}
    \caption{Sketch of the system in the ground state manifold of a QD including energy shifts and coupling to the light and magnetic fields. We only consider the ``$-$''-subsystem marked by color. The arrows indicate the spin configuration of the participating excitons.}
    \label{fig:levels}
\end{figure}
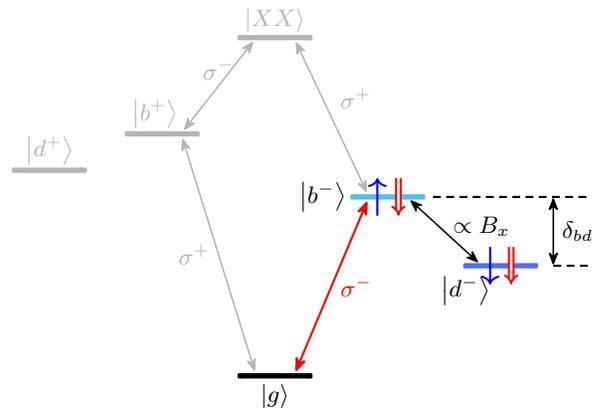

The energy levels and the most important couplings are sketched in Fig.~\ref{fig:levels}. In addition to the four exciton states $\ket{b^{\pm}}$ and $\ket{d^{\pm}}$, there exist the ground state $\ket{g}$ with no excitation and the biexciton state $\ket{XX}$. Due to the bright-dark splitting and the energy shifts induced by the $z$-component of the magnetic field, all single-exciton states are at different energies. The figure also depicts the optical selection rules for circularly polarized pulses. 

Assuming an excitation with only $\sigma^-$-polarized light, we can reduce our system to a three-level system consisting of  $\{\ket{g},\ket{b^-},\ket{d^-}\} \equiv \{\ket{g},\ket{b},\ket{d}\}$. While in principle the two bright excitons would be coupled via the fine-structure splitting, for typical splittings of a few tens of \si{\micro\electronvolt} \cite{Bounouar2018}, the conversion is much slower than our preparation scheme and gets further suppressed due to the magnetic field. As such, a preparation into the biexciton state is not feasible for a given circular polarization.

The Hamiltonian can then be written as 
\begin{equation}
    H = H_0\left(B_z\right) + H_{\text{flip}}(B_x)+H_{\text{light}}
\end{equation} 
with $H_0$ accounting for the energies of the above electronic states in the out-of-plane magnetic field $B_z$,  $H_{\text{flip}}$ for the spin flips induced by the in-plane magnetic field $B_x$ and $H_{\text{light}}$ for the coupling to the light field. For the three levels, $H_0(B_z)$ reads
\begin{align}
            H_0(B_z) &= 
            \hbar \omega_0\ket{b}\bra{b}  + \left(\hbar \omega_0 - \delta_{bd} \right) \ket{d}\bra{d},
\end{align}
where the energy of the ground state is set to zero, $\hbar\omega_0$ is the energy of the bright state and the dark state is shifted by $\delta_{bd}$. The energy splitting between bright and dark states is determined by the intrinsic bright-dark splitting $\delta_0$  and can be adjusted by the $z$-component of the magnetic field $B_z$, such that the effective splitting reads $\delta_{bd}(B_z) = \delta_0 - g_{\text{e},\,z}\,\mu_\text{B} \,B_z$. $\mu_\t{B}$ is the Bohr magneton. Because the two exciton states differ just by their electron spin component, here only the electron g-factor $g_{\text{e},\,z}$ enters. 

The in-plane magnetic field $B_x$ can result in a spin-flip of the electron, which couples bright and dark excitons
    \begin{equation}
        H_{\text{flip}}(B_x) = -\frac{J}{2} \ket{b}\bra{d} + \mathrm{h.c.}
    \end{equation}
via the coupling constant $J(B_x)=g_{\text{e},\,x}\,\mu_\text{B}\, B_x$. 

The interaction with the light field is treated in the dipole and rotating wave approximation reading
    \begin{equation}
        H_{\text{light}} = -\frac{\hbar}{2}\Omega\left(t\right)\ket{b}\bra{g} + \mathrm{h.c.} \,.
    \end{equation}
The electric field is given by $\Omega(t)$. We will consider two types of pulses: \begin{eqnarray}
    \Omega_{\t{P}}(t)&=&\frac{\theta}{\sqrt{2 \pi}\tau}e^{-t^2/(2\tau^2)}e^{-i\omega_\t{P} t}\\
    \Omega_\t{S}(t)&=& \frac{\Omega_0}{(1+e^{-\alpha\cdot t})(1+e^{-\alpha\cdot(\tau_{\text{S}}-t)})}e^{-i\omega_\t{S} t} \label{equ:Stark_Pulse}
\end{eqnarray}
On the one hand we consider short laser pulses with a Gaussian envelope which are characterized by their pulse area $\theta$ and $\tau=0.15$~ps. The carrier frequency of this laser pulse is set to be resonant with the transition frequency to the bright exciton $\omega_\t{P}=\omega_0$. In addition, we consider Stark pulses which have an essentially constant amplitude $\Omega_0$ during their length $\tau_{\text{S}}$. We assume that the Stark pulse has softened edges with a rise time of $1/\alpha=0.1$~ps. The carrier frequency of the Stark pulses is detuned from the transition frequency. To shift the eigenenergies of the system to lower values, we have to use a positive detuning, which is given by $\Delta=\omega_\t{S}-\omega_0$.

To compute the time evolution we use the density matrix formalism and numerically integrate the von-Neumann equation. Initially, we take the system to be in the ground state $\ket{g}$.

During the Stark pulse, we can write the Hamiltonian in matrix form in the basis $(\ket{g},\ket{b},\ket{d})$ as
\begin{align} \label{eq:H_rot}
    \tilde{H}=\begin{pmatrix} \hbar \Delta & -\frac{\hbar}{2}\Omega_ 0 &0 \\ -\frac{\hbar}{2}\Omega_0 & 0& -\frac{1}{2}J\\  0 & -\frac{1}{2}J& - \delta_{bd}
    \end{pmatrix} \,.
\end{align} 
Here we have chosen a frame, where the exciton energy becomes zero, while the ground state energy is given by the detuning. Note that this frame differs from typical rotating frames, but allows us an intuitive interpretation of the results. By diagonalization of this matrix we obtain the energy eigenstates of the system $E_i$, which we refer to as dressed states $\ket{i}$ with $i\in\{1,2,3\}$ in the following. While we perform the calculations in the bare state basis, the dressed states are useful for the interpretation of the dynamics. 

\subsection{Optical Stark effect}\label{subsec_stark_effect}
In a two-level system composed of the optically active states $\ket{g}$ and $\ket{b}$ , the action of the Stark pulse can be calculated by diagonalization of the corresponding Hamiltonian in the rotating frame
\begin{align}
    \tilde{H}=\begin{pmatrix} \hbar \Delta & -\frac{\hbar}{2}\Omega_0\\ -\frac{\hbar}{2}\Omega_0 & 0\\
    \end{pmatrix}\,.
\end{align}  
Note that we here set $J=0$ to focus on the optical Stark effect, such that we can omit the third state from Eq.~\eqref{eq:H_rot}. The diagonalization yields the eigenenergies
\begin{align}
    E_{\pm}(\Omega_0)=\frac{\hbar}{2}\Delta\pm \frac{\hbar}{2}\sqrt{\Delta^2+ \Omega_0^2}\, .
\end{align}
From this equation it is evident that applying a laser introduces an additional energetic splitting of the two eigenstates, which are now mixtures of the original ones. The Stark shift induced by the light field can be written as
\begin{align}
    \Delta E_{\text{Stark}}=&\,\,| E_{\pm}(\Omega_0)- E_{\pm}(0)| \,.
\end{align}
Even with the in-plane magnetic field ($J\neq0$), the influence of the third level $\ket{d}$ is negligible in most of the cases considered here. On the one hand, it is optically inactive and hence no direct coupling to the Stark pulse is possible. On the other hand, the dark state gains some optical activity, which makes it subject to the influence of the Stark pulse, but we assume in most cases that the coupling $J$ is much smaller than the bright-dark spitting $\delta_{bd}$, making this contribution very small.

\section{Preparation schemes} \label{sec:preparation}
Next, we will use pulse sequences to prepare different states in a GaAs QD. We set the coupling constant to $J=\SI{0.11}{meV}$, which for an electronic g-factor of $g_{\text{e},\,x}=-0.65$ \cite{bayer2000spectroscopic} corresponds to an in-plane magnetic field of $B_x=-3$~T. If not denoted otherwise, we use a bright-dark splitting of $\delta_{bd}=1.5$~meV, to illustrate our schemes. 

\subsection{Preparation of the dark exciton}\label{subsec:preparation_scheme}

\begin{figure}
    \input{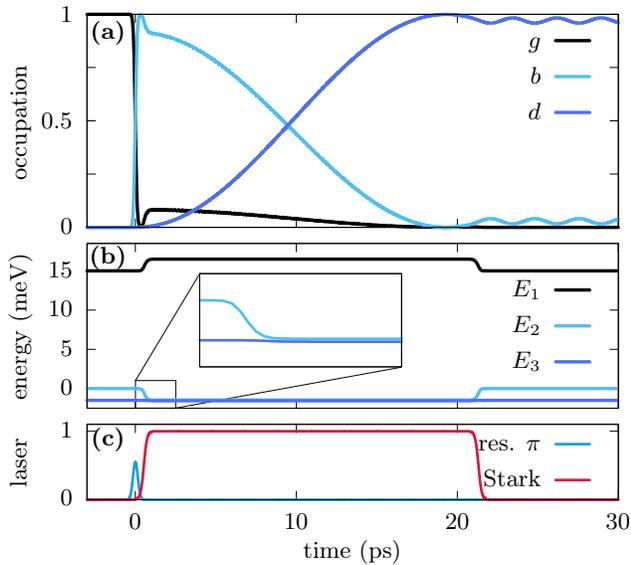}
    \caption{
    (a) Dynamics of the occupations $g$, $b$ and $d$ of the states in the three-level system, (b) evolution of the dressed-state energies $E_i$ and (c) sequence of the laser fields. Parameters for this example are $\hbar \Delta = 15$~meV, $\delta_{bd} = 1.5$~meV and $\hbar \Omega_0 = 9.95$~meV. 
    }
    \label{fig:dyn_ideal}
\end{figure}

We start by proposing a method for preparing the dark exciton state using the optical Stark effect, i.e., we aim for a high occupation of the dark exciton $\ket{d}$ after the application of the Stark pulse. In Fig.~\ref{fig:dyn_ideal} the dynamics of (a) the occupations $g$, $b$ and $d$ of the involved states, (b) the corresponding dressed-state energies and (c) the used pulse sequence are shown. 

In the beginning the system is in the ground state. By means of a $\pi$-pulse applied at $t=0$, which is resonant with the transition between the ground state and the bright exciton, the occupation switches completely to the bright exciton. After this, the off-resonant Stark pulse is switched on. The Stark pulse is a flat pulse that is positively detuned by $\hbar\Delta=15$~meV with respect to the transition between the ground state and the bright exciton. During the Stark pulse, the bright state population diminishes in favour of the dark state population which almost rises to unity. At $t\approx 21$~ps we turn off the Stark pulse and small oscillations between the bright and dark state occupation remain.

During the action of the Stark pulse the bright and dark exciton states are shifted into resonance. The behavior can also be understood by looking at the dressed-state energies as shown in Fig.~\ref{fig:dyn_ideal}(b). Without the Stark pulse $E_1$ corresponds to the ground state $\ket{g}$, $E_2$ to the bright state $\ket{b}$ and $E_3$ to the dark state $\ket{d}$. The Stark pulse now brings $E_2$ and $E_3$ close together and diabatic transitions between them take place. In order to preserve maximum dark state occupation, the Stark pulse is switched off when the occupation reaches its maximum, such that the eigenstates again shift out of resonance. Because the two excitons are still coupled via the magnetic field, the system continues with a small, off-resonant oscillation. 

\begin{figure}
    \hspace*{-0.8cm}\input{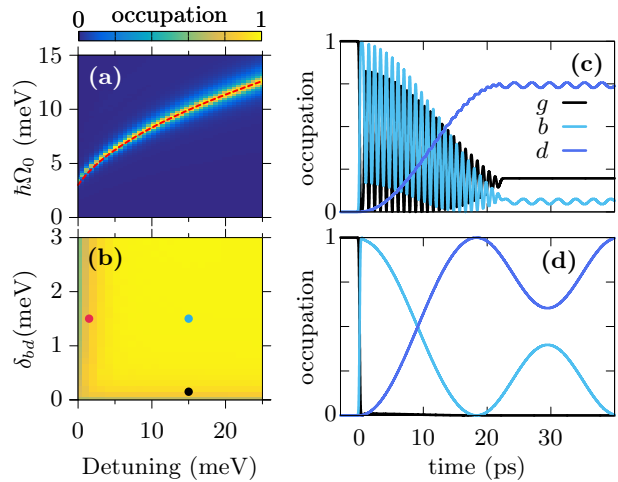}
    \caption{(a) Mean occupation of the dark exciton state after the action of the Stark pulse as function of detuning and strength of the Stark pulse $\hbar\Omega_0$ for a bright-dark splitting of $\delta_{bd}=\SI{1.5}{meV}$. The red dashed line indicates Eq.~\eqref{eq:resonance_condition}. (b) Final occupation as function of detuning and effective bright-dark splitting. The coloured dots indicate the parameters shown in the examples of the dynamics shown on the right (red and black) and in Fig.~\ref{fig:dyn_ideal} (blue). (c,d) Occupations $g$, $b$ and $d$ in the three-level system (c) for detuning $\hbar\Delta =1.5$~meV bright dark-splitting $\delta_{bd}=1.5$~meV and $\hbar \Omega_0=\SI{4.24}{meV}$ (red dot in (b)) and (d) $\hbar\Delta =15$~meV, $\delta_{bd}=0.15$~meV and $\hbar \Omega_0=\SI{3.01}{meV}$ (black dot in (b)).
    }\label{fig:accurancy}
\end{figure}

We now want to quantify the parameters for a Stark pulse, that leads to a high occupation of the dark exciton. For this, we first analyze the condition that brings the bright and dark exciton states into resonance. To this end, we vary the detuning and the amplitude of the Stark pulse and calculate the mean occupation of the dark exciton after the pulse as shown in Fig.~\ref{fig:accurancy}(a). We see that there is a clear condition for a high population of the dark exciton. This can be quantified by assuming that the energy shift induced by the optical Stark effect should be equal to the bright-dark splitting, leading to the resonance condition (for the case $\Delta>0$ considered in this work)
\begin{align} \label{eq:resonance_condition}
    \delta_{bd}\stackrel{!}{=}|\Delta E_{\text{Stark}}|=\left|\,{\frac{\hbar}{2}\Delta-\frac{\hbar}{2} \sqrt{\Delta^2 +\Omega_0^2}}\,\right|.
\end{align} 
From this, we derive for the mean occupation of the dark exciton
\begin{eqnarray}\label{eq:occ_cond}
    d_\text{{final}}&=&d_{\text{max}} - d_{\text{osc}} \nonumber \\
     &=& \frac{J^2}{J^2+\left(\delta_{bd}-\left|\Delta E_{\text{Stark}}\right|\right)^2}-\frac{1}{2}\frac{J^2}{J^2+\delta_{bd}^{2}}\,.
\end{eqnarray} 
Here, $d_{\text{max}}$ is the maximal occupation during the Stark pulse, which in the resonant case is $d_{\text{max}}=1$. Due to the switch-off and the residual oscillation this maximal occupation is reduced by the second term $d_{\text{osc}}$. This model yields a good approximation of the numerical results as can be seen by the red dashed line in Fig.~\ref{fig:accurancy}(a), marking the analytic position of $d_{\text{max}}$.

Furthermore, from the period of the oscillation between the bright and dark exciton state we can derive the length of the Stark pulse necessary for the state preparation as 
\begin{align}
    \tau_{\text{S}}= \frac{(2n+1)\pi\hbar}{\sqrt{J^2+ \left(\delta_{bd}-\left|\Delta E_{\text{Stark}}\right|\right)^2}}\, ,\qquad n\in\mathbb{N}_0\, .
\end{align}

Beside the pulse strength, also the dark-bright splitting is a crucial input parameter. Figure \ref{fig:accurancy}(b) shows the occupation of the dark exciton as a function of detuning and bright-dark splitting $\delta_{bd}$. In all these cases we choose the strength $\Omega_0$ of the Stark pulse according to the resonance condition Eq.~\eqref{eq:resonance_condition}. For most parameters, a near-unity occupation can be achieved. Only for small detunings and small bright-dark splittings, the preparation of the dark exciton is limited. In a GaAs QD the typical intrinsic bright-dark splitting is $\delta_0=\SI{0.25}{meV}$ \cite{Bayer2002}. To achieve the chosen bright-dark splitting of $\delta_{bd}=1.5$~meV, with an electronic g-factor of $g_{\text{e},z}=-0.8$ \cite{Bayer1999} this would refer to an out-of-plane magnetic field of $B_z=27$~T. For a more realistic value of $B_z=4$~T, we obtain $\delta_{bd}=0.43$~meV and an optimal value of $\hbar \Omega_0 = 5.15$~meV, which also results in an occupation close to unity with an average occupation of 0.95 [see also Fig.~\ref{fig:accurancy}(b)]

Two examples for non-optimal preparation are shown in Fig.~\ref{fig:accurancy}(c) and (d). In Fig.~\ref{fig:accurancy}(c) we consider the case where the detuning $\hbar \Delta = 1.5$~meV is too small to ensure a close to unity preparation. Because of the small detuning, the Stark pulse still induces a strong oscillation between ground and bright exciton state. This hinders the oscillation between bright and dark exciton, such that after the pulse only an occupation of about 0.75 is achieved and the ground state is strongly occupied. In Fig.~\ref{fig:accurancy}(d) the bright-dark splitting is set to $\delta_{bd}=0.15$~meV. Here we find that after the Stark pulse, we still see a strong oscillation between bright and dark exciton, hence the median occupation is clearly below one. This is an important aspect to the necessity of the application of a magnetic field with a $z$-component. Taking just a magnetic field with in-plane direction would result in a bright-dark splitting equal to the intrinsic splitting. In typical GaAs QDs this splitting is often too small to ensure a high-fidelity preparation.

\subsection{Preparation of superposition states}\label{subsec:superposition}
\begin{figure}[h]
    \centering
    {\input{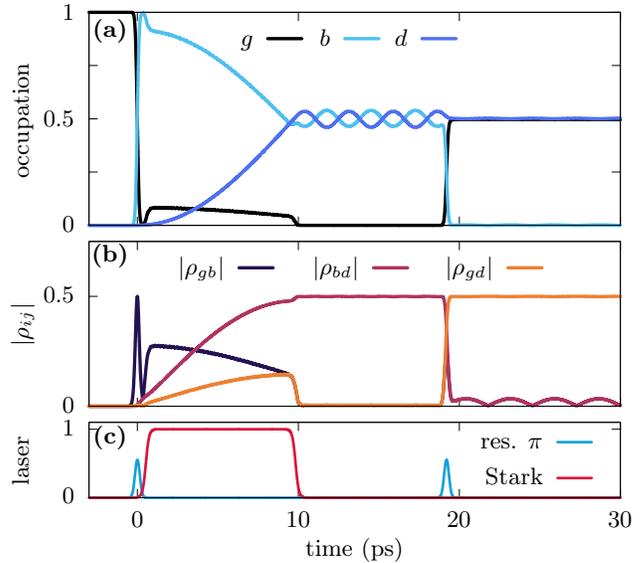}}
    \caption{Dynamics of (a) occupations $g$, $b$, $d$ of the states in the three-level system and (b) the coherences $\rho_{ij}$ with $i,j\in \{g,b,d\}$ under (c) the laser sequence. Same parameters as in Fig.~\ref{fig:dyn_ideal}. 
    }
    \label{fig:dyn_superposition}
\end{figure}

The possibility to switch between bright and dark states in a deterministic and precise fashion allows us to prepare different superposition states within the three-level system (cf.  Fig.~\ref{fig:levels}). An example of creating a superposition state is shown in Fig.~\ref{fig:dyn_superposition}, which displays (a) the occupations of the states alongside  (b) the coherences $\rho_{ij}$ for (c) a laser pulse sequence consisting of a $\pi$-pulse, a Stark pulse and a second $\pi$-pulse. 

The sequence starts the same way as the dark exciton preparation scheme, but this time we switch off the Stark pulse after half the time at $t\approx10$~ps, where the bright and dark states have a similar occupation of about $0.5$. As before, due to the in-plane magnetic field after the switch-off there is a small oscillation on the occupations. It is interesting to follow the dynamics of the coherences. Due to the switch-on of the Stark pulse, the coherence between ground and bright state $|\rho_{gb}|$ rises. Then, during the Stark pulse, we see that the coherence between bright and dark exciton $|\rho_{bd}|$ builds up. Also the coherence between ground and dark state $\rho_{gd}$ increases. This behaviour can be traced back to the composition of the involved dressed states $\ket{2}$ and $\ket{3}$, which are a mixture of all bare states. When switching off the Stark pulse both $|\rho_{gb}|$ and $|\rho_{gd}|$ go to zero and we obtain the superposition
\begin{equation}
    \ket{\psi}_{bd} =\frac{1}{\sqrt{2}}\left( \ket{b} + e^{-i\omega_{bd}t}e^{i\varphi_{bd}}\ket{d}  \right) \,.
\end{equation}
The coherence between $\ket{b}$ and $\ket{d}$ oscillates with the respective energy difference , i.e., with $\omega_{bd}=\delta_{bd}/\hbar$. During the protocol the coherence gains an additional phase $\varphi_{bd}=-\pi/2$.  After some waiting time, we apply a second $\pi$-pulse to de-excite the bright exciton to obtain the superposition state
\begin{equation}
    \ket{\psi}_{gd} =\frac{1}{\sqrt{2}}\left( \ket{g} + e^{i(\omega_{0}- \omega_{bd})t}e^{i\varphi_{gd}}\ket{d}  \right) \,.
\end{equation}
The coherence of $\ket{\psi}_{gd}$ oscillates with the energy difference between $\ket{g}$ and $\ket{d}$ with an additional phase $\varphi_{gd}$ gained during the action of the $\pi$-pulse. Note that if we were to wait for the spontaneous decay of the bright exciton, we would also expect to obtain the superposition state $\ket{\psi}_{gd}$, just with a different phase. The preparation of superposition states consisting of three or four energy levels are highly interesting for the generation of photonic cluster states \cite{lindner2009proposal,schwartz2016deterministic}. In our case a superposition of photons in different time bins could be created by successive de-excitation of the dark exciton by Stark pulses inducing only partial back-conversion to the bright exciton. 

\section{Influence of phonons}\label{subsec:phonons}
In QDs, phonons can strongly modify the state preparation schemes compared to the atomic case \cite{Reiter2014,luker2019review, Ramsay2010a, Reiter2019}. In the following we therefore take into account the coupling to longitudinal acoustic (LA) phonons, known to be the most important source of decoherence at low temperatures. 

\subsection{Exciton-phonon interaction}\label{subsec:theory_phonons}
The coupling to phonons is treated in the standard way via the Hamiltonian
\begin{align}
H_\text{ph}=&\, \hbar \sum_{\vec{q}} \omega_{\vec{q}} b_{\vec{q}}^{\dagger} b_{\vec{q}}^{}\\\nonumber
&\,+\hbar \left( \ket{b}\bra{b} +\ket{d}\bra{d}\right) \sum_{\vec{q}} \left(g_{\vec{q}} b_{\vec{q}} + g_{\vec{q}}^* b_{\vec{q}}^{\dagger} \right)\, ,
\end{align}
where $b_{\vec{q}}^{\dagger}$ ($b_{\vec{q}}$) is the bosonic creation (annihilation) operator for phonons with energy $\hbar\omega_{\vec{q}}$ that are coupled to the QD exciton states by $g_{\vec{q}}$. We assume a linear dispersion for the LA phonons and typical GaAs parameters \cite{luker2017phonon} at $T=1$~K for a QD radius of $3\,$nm.

When included in the equations of motion, the exciton-phonon interaction leads to the well-known infinite hierarchy of equations of motions, which we here truncate using a fourth-order correlation expansion formalism \cite{Kruegel2005}. It has been shown that for standard GaAs-QD coupling parameters at not too high temperatures, this method yields results coinciding with those given by numerically exact approaches such as the path integral formalism \cite{Vagov2011,Glaessl2011,vagov2011dynamics} and a good agreement with experiment \cite{kaldewey2017demonstrating}. 

Most importantly, the correlation expansion includes renormalization effects due to phonons as well as relaxation processes between the dressed states. It further describes effects of non-Markovian dynamics \cite{Reiter2019,carmele2019non}. The phonon renormalization of the excitonic energies is known as polaron shift and is given by
\begin{equation}
    \hbar \omega_0 \rightarrow \hbar \omega_0 - \hbar \Omega_{\text{pol}} = \hbar \omega_0 - \hbar \sum_{\vec{q}}\frac{|g_\vec{q}|^2}{\omega_{\vec{q}}}\, .
\end{equation}
It  is a result of the formation of a polaron quasiparticle. 
Therefore, in order to match the resonance condition, the polaron shift $\hbar\Omega_{\textrm{pol}}$ has to be taken into account. Because the phonons also change the Rabi frequency between ground and bright exciton state \cite{Kruegel2005,Ramsay2010b,Seidelmann2019} also the pulse areas have to be readjusted.
\subsection{Impact on preparation protocols}
\begin{figure}[h]
\centering
    \hspace*{-0.8cm}\input{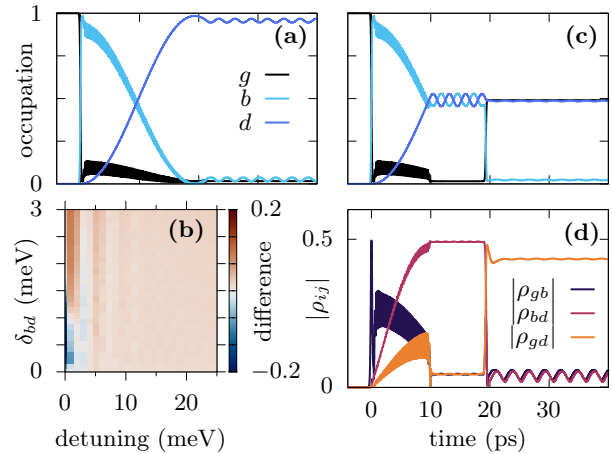}
    \caption{Dynamics of the occupation including the exciton-phonon interaction for $T=\SI{1}{K}$ for (a) the dark state preparation scheme [cf. Fig.~\ref{fig:dyn_ideal}] and (c) the preparation of the superposition [cf. Fig.~\ref{fig:dyn_superposition}] . (d) Dynamics of the coherences including the exciton-phonon interaction for the preparation into the superposition state in (c). (b) Difference of the final occupation without phonons [cf. Fig.~\ref{fig:accurancy}(b)] to the case including phonons as a function of detuning and bright dark-splitting. Note that a blue hue indicates a phonon enhancement of the preparation fidelity.
    }    
    \label{fig:phonons}
\end{figure}

To assess the impact of the phonon environment on our preparation protocol, we repeat the same protocols as in Sec.~\ref{sec:preparation} now accounting for the exciton-phonon interaction by using adjusted energies and pulse areas. Figure \ref{fig:phonons}(a) shows the occupation of the involved states for the dark state preparation for $T=1$~K. Similar to Fig.~\ref{fig:dyn_ideal} we find that after the pulses the dark exciton occupation is again close to unity. The marginal influence of the phonons can be understood by considering the relaxation processes induced by phonons in the dressed state basis [see Fig.~\ref{fig:dyn_ideal}(b)]. The dynamics start initially in $\ket{2}$, i.e., the middle dressed state, with diabatic transitions occurring to the lower dressed state. At low temperatures, the phonons can only lead to a transition from higher to lower lying dressed states due to phonon emission \cite{Lueker2012}, while phonon absorption is highly unlikely. However, due to the mixing coefficients and the large energy difference between the states, even the former transitions are highly unlikely. Hence, the phonons do not disturb the preparation scheme. The main difference in the case with phonons is a high-frequency oscillation with a small amplitude on top of the ground and bright exciton occupation as observed in Fig.~\ref{fig:phonons}(a). The origin of these oscillations is the phonon damping during the $\pi$-pulse, which results in a non-unity preparation of the bright exciton. Accordingly, also a superposition of the dressed states $\ket{1}$ and $\ket{3}$ occurs, where the high energy splitting between them corresponds to the frequency of the fast oscillations. 

Also, phonons do not noticeably disturb the preparation of the superposition state. The corresponding calculation with phonons at $T=1$~K is shown in Fig.~\ref{fig:phonons}(c), which shows similar occupations to the phonon-free case [cf. Fig.~\ref{fig:dyn_superposition}]. As the phonon environment is well known to dampen the Rabi rotations in the two-level system \cite{Machnikowski2004,Ramsay2010a,Kruegel2005}, it is interesting to analyze whether they affect the coherences. However, as demonstrated in Fig.~\ref{fig:phonons}(d) also the coherences are mostly unaffected by the phonons. 

To summarize the influence of phonons, Fig.~\ref{fig:phonons}(b) shows the difference of the final dark state occupation between calculations without and with phonons. For a wide range of excitation conditions, quantified by the laser detuning, and a variety of dark-bright splittings, the difference is in the percentage range and thus very small. Only at small detunings, where the amplitude of these oscillations is high, the damping by phonons is especially effective, because here strong oscillations between ground and bright exciton state take place, which are highly affected by phonons. Therefore, the final occupation of the dark state is also affected, since our protocol relies on the bright state's occupation as an intermediary. It is noteworthy that there are not only regions where the final dark state occupation is dampened by phonons, but also regions of phonon enhancement can be found [blue-hued areas in Fig.~\ref{fig:phonons}(b)]. In this region the energetic splitting between the eigenenergies $E_1$ and $E_2$ is in the range of the phonon spectral density, which has its maximum at $\SI{2}{meV}$ for our parameter set. This leads to phonon-assisted transitions \cite{luker2017phonon} from the ground state to the bright exciton state and therefore supports the preparation of the dark exciton.

For very high detunings, the splitting between the dressed states $\ket{1}$ and $\ket{2}$, which have mainly ground state and exciton characteristics, can surpass the phonon spectral density and phonon processes can be completely suppressed \cite{kaldewey2017demonstrating}. In these regions, our preparation protocol becomes also stable against phonon influence at higher temperatures. 

\section{Experimental feasibility of Stark pulses}
The shown method relies on the application of two different kinds of laser pulses, a resonant $\pi$-pulse followed by a spectrally detuned Stark pulse. Exciton preparation via resonant $\pi$-pulses has already been demonstrated many times \cite{stievater2001rabi,ramsay2010review} and provides no new experimental challenge since one can use commercially available pulsed laser systems that can provide pulse durations of the order of \SI{100}{\femto\second}. However, generating a laser pulse of the form presented in Eq.~\eqref{equ:Stark_Pulse} requires more attention. 

\begin{figure}[t]
    \centering
    \hspace*{-0.2cm}\input{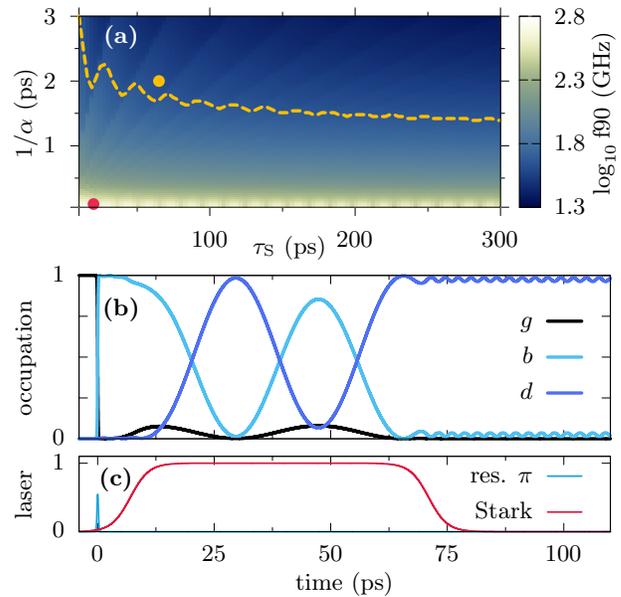}
    \caption{(a) Classification of the electric driving signal as a function of $1/\alpha$ and $\tau_\text{S}$. Colors indicate the frequency where \SI{90}{\percent} of the electrical driving signal are present (f90). Orange dashed line: 45 GHz line for modern EOM systems. The dots indicate the value used for the dark state preparation (red: Fig.~\ref{fig:dyn_ideal}; orange: this figure). (b) Dynamics of the occupation of the three-level system without phonons for the pulse sequence shown in (c).  Parameters for this example are $\hbar \Delta = 15$~meV, $\delta_{bd} = 1.5$~meV and $\hbar \Omega_0 = 10.1$~meV. 
    \label{fig:Stark_experimental}
    }
\end{figure}

One straightforward way of generating such pulses is to utilize fast electro-optical modulators (EOMs) and cut the desired pulses from a continuous wave (CW) laser source operating at the desired wavelength. A clear advantage of this method is that one can use the $\pi$-pulse and a sufficiently fast photo-diode as a trigger to start the driving signal for the EOM, which is much easier than synchronizing two pulsed laser sources. \\ 
Commercially available EOM systems nowadays support a bandwidth of up to \SI{45}{\giga\hertz} which restricts the experimentally possible values of $\alpha$ and $\tau_{\text{S}}$ in Eq.~\eqref{equ:Stark_Pulse}. Sweeping these parameters, calculating the respective electrical bandwidth and integrating to a point where \SI{90}{\percent} of the driving signal's frequency contributions are present, allows us to check for possible values for $\alpha$ and $\tau_\text{S}$ as shown in Fig.~\ref{fig:Stark_experimental}(a).   

As evident from Fig.~\ref{fig:Stark_experimental}(a) there exists a limit for the experimentally realizable values of $\alpha$ and $\tau_{\text{S}}$ using currently available EOMs, which is indicated by the orange dashed line. In particular, everything above this orange line can be realized with current EOMs. The red dot indicates the present parameters of the Stark pulse as used in Fig.~\ref{fig:dyn_ideal} with $1/\alpha=0.1$~ps and a length of about $20$~ps. We find that these parameters are strongly below what is currently achievable in experiments.

To achieve a feasible length of the Stark pulse we need to use much longer pulses of several tens of ps. This can be done in several ways. One possibility would be to reduce the in-plane magnetic field $B_x$ and accordingly lengthen the period of the oscillation between bright and dark exciton state. Another opportunity is to wait for more than one half period to switch off the Stark pulse. The second aspect is, that even for a longer pulse length, the rise time of the Stark pulse has to be increased, such that it is on the order of a few ps.
\\
To test if our proposed methods still work for such less favourable pulse parameters, Figs.~\ref{fig:Stark_experimental}(b) and (c) show the population dynamics and laser sequence for currently realizable experimental parameters: $1/\alpha = \SI{2}{\pico\second}$ and $\tau_\text{S} = \SI{64}{\pico\second}$ [orange dot in Fig.~\ref{fig:Stark_experimental}(a)]. We take the same magnetic field values as in Fig.~\ref{fig:dyn_ideal} and wait for one and a half oscillation to take place before switching off the pulse. In the case without phonons, the longer rise time and Stark pulse length also leads to a near unity population of the dark exciton state utilizing one and a half oscillations of the induced dark-bright splitting. These preparation times are still short compared to the lifetime of the bright exciton, which is on a time scale of several hundred of ps up to a ns \cite{syperek2012influence}, therefore we think that even when using longer pulses our proposed scheme is still feasible. 

\section{Conclusion}\label{sec:conclusion}
In summary, we have analyzed the usage of the optical Stark effect to prepare and control the dark state excitation for a QD in a tilted magnetic field. The Stark effect allows preparing the dark exciton with almost unity fidelity, even under the influence of phonons. By adjusting the length of the Stark pulse, we can prepare different superposition states including the dark exciton. Coherent control of the dark state occupation can be useful to enhance already existing or establish new preparation protocols in quantum dots to generate new light states like entangled or cluster states.

\acknowledgements
We thank the Austrian Science Fund FWF and the German Research Foundation DFG for support through the D-A-CH project "Advanced Entanglement from Quantum Dots" (DFG research fund number 428026575).
M.C. gratefully acknowledges support by the Studienstiftung des Deutschen Volkes.

\bibliography{bibliography,bib_Michael,bib_Doris}

\end{document}